\documentclass[a4paper, pra, twocolumn, superscriptaddress,showpacs]{revtex4}

\usepackage{amssymb}
\usepackage{amsmath}
\usepackage{color}
\usepackage{graphics, graphicx}
\usepackage{psfrag}
\usepackage{mathcomp}
\usepackage{subfigure}
\usepackage{color}
\usepackage{soul}

\newcommand{\mutilde}{{\tilde{\mu}}}
\newcommand{\Deltatilde}{{\tilde{\Delta}}}
\newcommand{\Lambdatilde}{{\tilde{\Lambda}}}
\begin{document}

\author{A. Privitera}
\affiliation{Democritos National Simulation Center, Consiglio Nazionale delle Ricerche, Istituto Officina dei Materiali (IOM) and Scuola Internazionale Superiore di Studi Avanzati (SISSA), Via Bonomea 265, 34136 Trieste, Italy}
\author{M.Capone} 
\affiliation{Democritos National Simulation Center, Consiglio Nazionale delle Ricerche, Istituto Officina dei Materiali (IOM) and Scuola Internazionale Superiore di Studi Avanzati (SISSA), Via Bonomea 265, 34136 Trieste, Italy}
\affiliation{Dipartimento di Fisica, Universit\`a di Roma ``La Sapienza'', Piazzale Aldo Moro 2, I-00185 Roma, Italy}

%opening
\title{Lattice approaches to dilute Fermi gases: Legacy of broken Galilean invariance}
\begin{abstract}
In the dilute limit, the properties of fermionic lattice models with short-range attractive interactions converge to those of a dilute Fermi gas in continuum space. We investigate this connection using mean-field and we show that the existence of a finite lattice spacing has consequences down to very small densities. In particular we show that the reduced translational invariance associated to the lattice periodicity has a pivotal role in the finite-density corrections to the universal zero-density limit. 
 For a parabolic dispersion with a sharp cut-off,  we provide an analytical expression for the leading-order corrections in the whole BCS-BEC crossover. These corrections, which stem only from the unavoidable cut-off, contribute to the leading-order corrections to the relevant observables. In a generic lattice we find a universal power-law behavior $n^{1/3}$ which leads to significant corrections already for small densities. Our results pose strong constraints on lattice extrapolations of dilute Fermi gas properties. 

\end{abstract}
\maketitle

 \section {Introduction}
One of the most attractive features of ultracold Fermi gases is the universal physics arising from their intrinsic dilution. A Fermi gas of volume density $\rho$ is dilute when $k_F R_0 \ll 1$, where $k_F=(3\pi^2\rho)^{1/3}$ is the Fermi momentum and $R_0$ is the range of the interaction. When this condition is realized the interactions are controlled only by the s-wave scattering length $a_s$ and the details of the potential become irrelevant. Therefore the physics is controlled by the inverse gas parameter $\eta=(k_Fa_s)^{-1}$ and a universal phase diagram can be drawn, regardless of the nature of the fermions involved \cite{varennalectures}. 
Moreover, Fano-Feshbach resonances \cite{fano-feshbach} allow to tune the interaction from weak- to strong-coupling realizing (i) The crossover from the Bardeen-Cooper-Schrieffer (BCS) superfluidity ($\eta \to -\infty$) to a Bose-Einstein Condensation (BEC) of preformed pairs ($\eta \to +\infty$)\cite{zwerger} and (ii) The unitary limit in which the scattering length diverges, and the system has no characteristic length scale other than $k_F^{-1}$ leading to an extra universality which attracted a remarkable interest \cite{unitary}.

In a different framework, the physics of superconductors beyond the BCS regime, including the high-Tc materials, has triggered the investigation of lattice models with attractive interactions. The typical electronic densities in these materials are far from the dilute limit, being instead closer to the so-called half-filling regime in which the number of fermions equals the number of lattice sites. However, a BCS-BEC crossover takes place also in this regime \cite{xover-lattice}  even if no universality is expected.

In this work we focus on the connection between these two different realizations of the BCS-BEC crossover. The main effect of the lattice is to lift the Galilean translational invariance introducing a characteristic length scale, the distance between two lattice sites, or lattice spacing. The reduced translational invariance makes the fermionic spectrum bounded from above or, in other words, introduces an ultraviolet {\it energy cut-off}.

When the number of fermions is of the same order of the number of sites the underlying lattice influences the properties of the fermionic gas substantially. Reducing the density, the lattice becomes less and less relevant, and in the dilute limit the properties of lattice fermions are expected to converge to  
the universal features of dilute gases.
The evolution of the properties of a lattice model by reducing the density is an interesting topic {\it{per se}}, also in light of the possibility to study the BCS-BEC crossover  in optical lattices at various particle densities \cite{schneider}, but it can also be used to benchmark and test theoretical studies in which the properties of dilute gases are obtained through a low-density extrapolation of lattice models, as done in recent lattice Quantum Monte Carlo (QMC) approaches to the unitary Fermi gas \cite{bulgac1,bulgac2,bulgac3,burovski1,burovski2,burovski3,goulko,lee1, lee2,drut,carlson2011}. 

In a previous work \cite{ufg_old}, we used static and dynamical mean-field theory \cite{DMFTreview} at the unitary limit. We proved that finite-density corrections to the dilute unitary limit can be large even at small densities and that the dependence on the specific lattice is hardly washed out at any finite density. Therefore a safe extrapolation to the dilute limit is in general extremely difficult.

In this work we extend the analysis in the whole BCS-BEC crossover and we focus on the origin of the finite-density corrections.  In order to gain some analytical understanding, we use a static mean-field theory, which allows to treat on the same footing the lattice and continuum cases. The simplicity of the approach helps us to identify  the origin of the leading density-corrections to the dilute limit  together with their functional dependence on the inverse gas parameter $\eta$ in the whole crossover. 
 
The paper is organized as follows: In Sec. \ref{sec:mapping}, we define the connection between lattice and continuum models in terms of the relevant variables in the dilute regime. Then we introduce in Sec. \ref{sec:mf} the mean-field approach and we recast it
in terms of these variables. Results for the unitary system and in the whole crossover are shown in Sec. \ref{sec:results}. Finally we draw the concluding remarks in Sec. \ref{sec:conclusions}, together with the consequences of our findings over lattice extrapolation of the BCS-BEC crossover in dilute gases.

\section{The lattice path to the dilute limit}
\label{sec:mapping}

In this section we establish formally the connection between lattice models and dilute gases in continuum space.
First we identify the lattice counterpart of the inverse gas parameter $\eta$, then we discuss the role of the specific dispersion of the lattice and of the ultraviolet cut-off. Finally we cast some general considerations in the form of a phase diagram in the relevant parameter space.

\subsection{Interaction: the scattering length}
As briefly discussed in the Introduction, for a dilute Fermi gas interacting via a short-range potential  the only 
relevant parameter to describe the interaction is the s-wave scattering length $a_s$ and the properties of the 
system are defined by the inverse gas parameter $\eta = (k_Fa_s)^{-1}$. In this subsection we define the lattice 
versions of $a_s$ and $\eta$.

We model the lattice system with an attractive Hubbard model
\begin{equation}
 {\mathcal H} = -t \sum_{<ij>\sigma} c_{i\sigma}^{\dagger} c_{j\sigma} 
+ U\sum_{i} n_{i\uparrow} n_{i\downarrow} -  \mu \sum_{i\sigma} n_{i\sigma}
\label{HM}
\end{equation}
where $t$ is the hopping parameter, $U<0$ is the onsite attraction and $\mu$ is the chemical potential
which fixes the density of particles per lattice site $n$ in a grandcanonical ensemble. 
 
Since the scattering length $a_s$ is a two-body property, it can be evaluated from the sole knowledge of the interparticle potential. $a_s$ is proportional to the limit of vanishing energy and quasi-momentum of the 
the two-body vertex function in the vacuum $\Gamma_{2-body}^{latt}$ \cite{burovski2}, i.e.  
\begin{equation}
\label{as}
 a_s^{latt}= \frac{m l^3}{4\pi\hbar^2} \lim_{\omega ,\vec{q} \to (0,\vec{0})} \Gamma^{latt}_{2-body}(\omega ,\vec{q})
\end{equation}   
 where $m$ is the bare effective mass of the fermions in the lattice at hand and $l$ is the lattice spacing.  

For a Hubbard model 
$\Gamma_{2-body}^{latt}(\omega ,\vec{q})= [U^{-1}-\Pi^{latt}(\omega ,\vec{q})]^{-1}$, where $\Pi^{latt}$ is the polarization bubble
\begin{equation}
\label{pilatt}
\Pi^{latt}(\omega,\vec{q})=-\int_{FBZ} \left( \frac{l}{2\pi} \right)^d d \vec{k} \ \frac{1}{\omega  - \epsilon_{\frac{\vec{q}}{2}+\vec{k}}
- \epsilon_{\frac{\vec{q}}{2}-\vec{k}}+i 0^+},
\end{equation}
the momentum integral extends over the first Brillouin zone and the zero energy is chosen to have  non-negative energies and $\epsilon_{\vec{0}} = 0$.

(\ref{as}) and (\ref{pilatt}) determine the dependence of $a_s^{latt}$ on $U$
\begin{equation} 
a_s^{latt}(U) = \frac{m l^3}{4 \pi \hbar^{2}}\frac{1}{U^{-1}-{U_c}^{-1}}.
\label{scatteringlength}
\end{equation}
where $U^c$ is defined by $[\Pi^{latt} (0,\vec{0})]^{-1}\equiv U_c$ and, using $\epsilon_{\vec{k}}=\epsilon_{-\vec{k}}$,
\begin{equation}
\Pi^{latt} (0,\vec{0})= -\int_{FBZ}\left( \frac{l}{2\pi} \right)^d d \vec{k} \ \frac{1}{ 2 \epsilon_{\vec{k}}}=
-\int_0^\Lambda d \epsilon \frac{D_{latt}(\epsilon)}{2\epsilon}.
\label{bubble} 
\end{equation}
In the expression above we emphasized that the lattice density of states (DOS) $D_{latt}$ is finite only below a cut-off   $\Lambda$. The scattering length therefore diverges at $U_c$ which defines the unitary limit on the lattice. It is worth to notice that for $|U| \to \infty$ $a_s$ converges to a finite value of the order of the lattice spacing $l$. Hence even for divergent interactions the finite lattice spacing implies a non-zero size of the Cooper pairs.

In a lattice model it is natural to describe the system is terms of the number of particles per lattice site $n$ (also called filling factor). $n/l^3$ is the number of particles per unit volume and we can define the Fermi energy $E_F$ and the Fermi momentum $k_F$ of a non-interacting homogeneous gas with the same density and mass $m$ as
\begin{equation}
 E_F=\frac{\hbar ^2 k_F^2}{2m}    \ \ \ \  k_F = \left[ 3 \pi^2 \left( \frac{n}{l^3} \right) \right]^{\frac{1}{3}}.
\label{fermi_energy}
\end{equation} 

(\ref{scatteringlength}) and (\ref{fermi_energy}) define the lattice analogue of the inverse gas parameter $\eta=(k_Fa_s^{latt})^{-1}$. In the following we will study the attractive Hubbard model along lines with constant $\eta$ which span the whole BCS-BEC crossover.

\subsection{Lattice dispersion and the role of the cut-off}
\label{cut-off}
As outlined in the introduction, an unavoidable consequence of the broken Galilean invariance is the existence, in any finite dimension, of an ultraviolet cut-off $\Lambda$ in the dispersion. This corresponds to the finite bandwidth of lattice systems, as opposed to the unbounded parabolic dispersion of particles in free space. We will show that, even if the cut-off occurs at energies much larger than the Fermi energies corresponding to low-density ($E_F \propto n^{2/3}$), it contributes to the leading-order corrections in the density to the universal dilute limit.

\begin{center}
\begin{table}[tb]
\begin{tabular}{|c|c|c|c|c|c|c|}
\hline
DOS & Cut-off & $|U_c|$ & $R_e$ & $a_{\mu}$ & $a_{\Delta}$ & $a_{\xi}$ \\ \hline
% Cubic & 1.1 & 0.73224  & 0.3056\\
SC & 10.8103 & 5.4051 &  0.4301& -0.040 & -0.028 & -0.034 \\ \hline
% $LG_1$ & 7.5963 &  5.0642& 0.3267& -0.063 & -0.021& -0.052 \\ \hline  FIT RESULTS 
$LG_1$ & 7.5963 &  5.0642& 0.3267& -0.064 & -0.020 & -0.052 \\ \hline
% analytics
$LG_2$ & 14.8044 &  5.1444& 0.3369& -0.062 & -0.022& -0.050 \\ \hline
% Bulgac & 0.9129  & 1,176119 & ?\\
% Russi &  1.82596802686378 & 0.821974428572 & ?\\
\end{tabular}
\caption{DOS summary $(\hbar=l=m=1)$}
\label{tab:summary}
\end{table}
\end{center}

In order to properly reproduce the dilute limit of fermion gas  with mass $m$ in three-dimensional
homogeneous space we have to consider a class of lattice dispersions which share the same low-energy behavior of the free DOS $D_{free}(\epsilon)=\frac{(2m)^{3/2}}{4\pi^2\hbar^3}\sqrt{\epsilon}$.
This implies $D_{latt}(\epsilon) \propto \sqrt{\epsilon}$ (or $\epsilon_k \propto k^2$). 

\begin{figure}[!htbp]
\begin{center}
\includegraphics[scale=0.375]{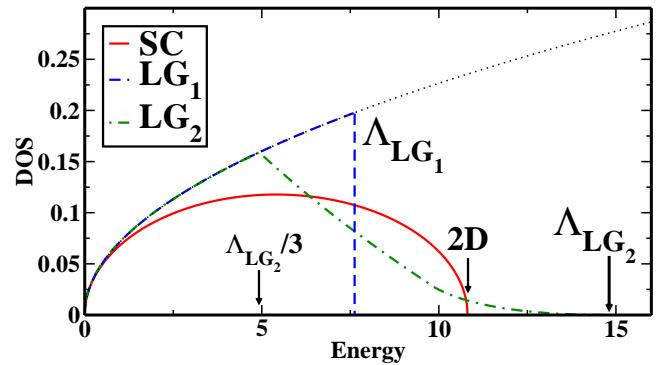}
\caption{(Color Online) Comparison between the different lattice DOS under investigation: Semicircular (full red line), Lattice Gas 1 
(dashed blue line), and Lattice Gas 2 (dashed-dotted green line). For reference the DOS per lattice site $l^3 D_{free}$ for free fermions in three-dimensional space is shown as a dotted black line. We set $\hbar=l=m=1$.}
\label{fig:dos}
\end{center}
\end{figure}

In order to identify how the lattice and continuum physics connect, we study the following three models (shown in Fig. \ref{fig:dos}), all sharing the correct low-energy behavior, but substantially different at high-energy
\begin{itemize}

\item{A semicircular DOS
\begin{equation}
\label{scdos}
D_{SC}(\epsilon)=\frac{2}{\pi D} \sqrt{\frac{\epsilon}{D}\left(2-\frac{\epsilon}{D}\right)}  
\end{equation}
where $D$ is the half-bandwidth 
and $\Lambda = 2D$ is the cut-off and $D=(4\pi)^{2/3}\frac{\hbar^2}{ml^2}$.}
\item{A first {\it lattice gas} model ($LG_{1}$) with a sharp cut-off, namely a system in which the DOS is that of free fermions up to a sharp energy cut-off $\Lambda_{LG_1}$ which enforces the presence of the lattice and ensures the correct normalization of the DOS per lattice site. Above the cut-off  the DOS is identically zero.}
\item{A second {\it lattice gas} model ($LG_{2}$) with a smoother cut-off. The dispersion coincides again with that of free fermions, but the lattice periodicity is enforced restricting the momenta to the first  Brillouin zone of a cubic lattice (${\bf{k}} \in [-\pi/l,\pi/l)^3$). In this way the DOS coincides with the free one only up to $\Lambda_{LG_2}/3$, then smoothly decreases and vanishes at $\Lambda_{LG_2}$.}
\end{itemize}

By comparing these three models, we can clarify the roles of the cut-off $\Lambda$ and the effect of the different behavior of the DOS'S at the various energies.

In order to properly compare the different lattices, we evaluated according to Eq. (\ref{bubble}) the critical value $U_c$ for each lattice. Results are summarized in Tab. \ref{tab:summary}, together with the value of the cut-off and other quantities defined in the following.

\vspace{0.125cm}

We do not study the cubic lattice, which has been shown in Ref. \cite{ufg_old} to introduce larger finite-density effects than the models defined above.

\subsection{Approach to the dilute limit}

\begin{figure}[!ht]
\begin{center}
\includegraphics[scale=0.32]{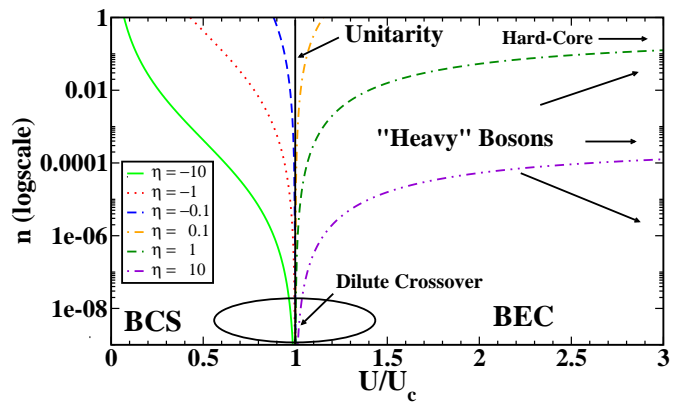}
\caption{(Color Online) Trajectories with constant $\eta $ in the $(|U|,n)$ plane of an attractive Hubbard model 
 for $\eta =-10,-1,-0.1$ (BCS side), $\eta =0$ (Unitarity) and $\eta =0.1,1,10$ (BEC side). 
The results shown concern the semicircular DOS (see Sec. \ref{sec:mapping}) but a similar picture 
is expected in the other cases.}
\label{fig:eta_vs_U}
\end{center}
\end{figure}

 Once the mapping $(U,n)\to (a_s,k_F)$ is established via Eqs. (\ref{scatteringlength}) and (\ref{fermi_energy}), 
we can consider the parameter space of the lattice model at $T=0$ in terms of the inverse gas parameter $\eta$, which controls the physics in the dilute limit. By expressing $\eta$ as a function of the coupling $U$ and of the density $n$ and inverting this relation, we can calculate the trajectories for fixed $\eta$ in the $(|U|,n)$ parameter plane of the attractive Hubbard model, i.e. 
\begin{equation} 
\frac{U_\eta(n)}{U_c}=\frac{1}{1-\alpha \eta n^{1/3}}
\label{U_eta}
\end{equation}
where $\alpha=\sqrt[3]{\frac{3}{64\pi}}\left(\frac{ml^2|U_c|}{\hbar^2}\right)$ is a dimensionless constant depending on the lattice at hand. In Fig. \ref{fig:eta_vs_U} we plotted these trajectories for several values of the inverse gas parameter $\eta$ in the BCS-BEC crossover for a semicircular DOS. These trajectories give a pictorial representation of how the physics of a lattice model at finite density evolves towards a continuum system in the dilute limit. 

It should be clear, however, that this diagram is not meant to provide information about the density where we expect to get rid of non-universal lattice features and  access the universal properties of dilute gases. This is particularly evident for the trajectory at unitarity $\eta=0$, where, even though the correct value of the interaction is $U_c$ for any density, the universal properties of the unitary Fermi gas can be recovered only at small density.

As shown in Fig. (\ref{fig:eta_vs_U}), the trajectories converge at low-density towards the critical value $U_c$. While at large densities the crossover is wide in $U$, in the dilute region everything happens in a very narrow region around $U_c$, and the values far from $U_c$ are safely in the BCS and BEC regimes, respectively.

We expect that the convergence to the dilute regime is slower on the BEC side, where the property of the superfluid state are more sensitive to the presence of the lattice. Indeed, in the strong coupling regime for the lattice model $\frac{|U|}{t} \gg 1$ the size of the superfluid pairs is bounded from below by the lattice spacing. In this regime the fermions form strongly bound pairs with a large effective mass $M_B \propto \frac{|U|}{t} \gg m$ (heavy bosons region in Fig. \ref{fig:eta_vs_U}). Despite of the hard-core constraint arising from the Pauli principle being irrelevant at low-density, even a single pair can only move through virtual ionization due to the presence of the lattice. This implies that the reduced translational symmetry of the lattice is relevant at any finite density in this regime.  

It is useful to notice that, due to the fact that the scattering length cannot be arbitrarily reduced in the lattice, the trajectories for $\eta >0$ have an horizontal asymptote for large $\vert U\vert$, implying that large positive values of $\eta$ cannot be accessed in the lattice above a given density.

On the other hand, the BCS regime is not expected to be {\it qualitatively} dependent on the density allowing for a smoother extrapolation to the dilute limit. This expectation is confirmed by the results that we present in the Section \ref{sec:results}.

\section{Mean-field approach}
\label{sec:mf}

After the general considerations above, we analyze quantitatively the convergence 
of the lattice models to a dilute Fermi gas in continuum space by means of a static mean-field approach.
As shown by Leggett \cite{leggett} for the dilute gas and by Nozi\`eres and  Schmitt-Rink for the lattice model \cite{NSR}, the mean-field approach provides a reasonable description of the ground state properties through the whole BCS-BEC crossover, being exact both in the BCS and BEC limit and interpolating between them at intermediate coupling. Within this simple approach we can deal with continuum and lattice system on equal footing without introducing any bias. Another advantage over more accurate methods is that it  can be 	implemented directly in the thermodynamic limit, getting rid of finite-size effects, and in the grand canonical ensemble, which allows to address arbitrarily small densities.
 
We implement a mean-field approach for the attractive Hubbard model and we follow trajectories of fixed $\eta$ in the parameter space. For vanishing density we expect to recover the results of Leggett's mean-field theory for dilute Fermi gases \cite{leggett}. 

The static mean-field equations for the Hubbard model in the superfluid phase are easily obtained by considering the Hamiltonian (\ref{HM}) and decoupling the interaction term both in the particle-hole and in the
particle-particle channel and introducing a gap parameter $\Delta=U \langle \hat{c}_{i,\uparrow} \hat{c}_{i,\downarrow}\rangle$. The self-consistency equations read
\begin{eqnarray}
n &=&\int d\epsilon\  D_{latt} (\epsilon)  
\left( 1-\frac{\zeta(\epsilon)}{\gamma(\epsilon)} \right) \label{mflattice1} \\
\frac{1}{|U|}  &=& \frac{1}{2} \int d\epsilon\ \ \frac{ D_{latt} (\epsilon)}{\gamma(\epsilon)} \label{mflattice2}  
\end{eqnarray}
where $\zeta(\epsilon)=(\epsilon-\mu^\prime)$, $\gamma(\epsilon)=\sqrt{(\epsilon-\mu^\prime)^2+\Delta^2}$ 
and the Hartree term is included in the shifted chemical potential $\mu^\prime=\mu-(Un/2)$. 
The ground state energy per lattice site $E$ is given by 
\begin{equation}
E =\int d\epsilon\  D_{latt} (\epsilon) 
 \left[ \zeta(\epsilon) - \gamma(\epsilon) +\mu  
 \left(1-\frac{\zeta(\epsilon)}{\gamma(\epsilon)} \right)
 \right] - E_0 
\label{MFenergy}
\end{equation}
where $E_0 = Un^2/4 +\Delta^2/U$. The self-consistency equations (\ref{mflattice1}) and (\ref{mflattice2}) have been already studied in \cite{NSR} as a function of the interaction $U$ at fixed density. The focus of this work is, however, the evolution towards low density and the connection for $n\to0$ to the Leggett mean-field approach in the continuum \cite{leggett}. In this way we  can use them to gain insight on the finite-density corrections to the dilute limit in a lattice. 

By solving numerically the mean-field equations for the different lattice models under consideration we can assess that our method reproduces Leggett's results as derived in Ref. \onlinecite{strinati_analytic} in the whole BCS-BEC crossover and, most importantly, we quantify the deviations from the dilute limit as a function of the density and interaction, together with their dependence on the specific lattice at hand. 

By inverting the expressions of the Fermi energy $E_F$ in Eq. (\ref{fermi_energy}) and of the inverse
gas parameter $\eta(U,n)$ in Eq. (\ref{U_eta}), we can recast Eqs. (\ref{mflattice1},\ref{mflattice2}) in terms of the relevant variables for the dilute limit, obtaining an equivalent set of equations for the renormalized gap parameter $\tilde{\Delta}=\frac{\Delta}{E_F}$ and the renormalized chemical potential $\tilde{\mu}=\frac{\mu}{E_F}$. 

After some algebra one obtains
\begin{eqnarray}
 \frac{4}{3}=&&\int_{0}^{\Lambda/E_F}\hspace{-0.5cm}d x\  \tilde{D}(x) \left[ 1 - \frac{(x-\tilde{\mu^\prime})}
{\sqrt{(x-\tilde{\mu^{\prime}})^2 + \tilde{\Delta}^2}}\right] \label{mf1} \\
 \eta =\frac{1}{\pi}&&\int_{0}^{\Lambda/E_F}\hspace{-0.5cm} d x\  \tilde{D}(x)
\hspace{-0.1cm} \left [ \frac{1}{x} - \frac{1}{\sqrt{(x-\tilde{\mu^\prime})^2 + \tilde{\Delta}^2}} \right] \label{mf2}
\end{eqnarray}
where $x=\epsilon/E_F$,  $\tilde{D}(x)=\frac{D_{latt}(E_F x)}{l^3 D_{free}(E_F)}$,  $\tilde{\mu^\prime} =  \tilde{\mu} - \frac{n U_\eta(n)}{2 E_F}$
is the renormalized chemical potential including the Hartree contribution, $\Lambda$ is the energy cut-off which derives from the broken translational invariance and corresponds to the lattice bandwidth in the solid state framework. We remark that we are not assuming that the density $n$ is small at this step, but we just 
reformulated the mean-field equations in a lattice in terms of the relevant variables in the dilute limit.   

It is easy to verify that for $n\to0$, i.e. $E_F \to 0$, and fixed $\eta$, our mean-field equations reduce asymptotically to Leggett mean-field equations \cite{leggett}. In this limit both the gap and the chemical
potential are proportional to $E_F$ through coefficients which are only function of $\eta$. On the other hand, as we shall discuss in the following, for any finite density $\tilde{\mu}$ and  $\tilde{\Delta}$ retain an additional dependence on the density according to Eqs. (\ref{mf1},\ref{mf2}).

The corrections to Leggett theory arise from two different sources: (i) the different DOS of the lattice models and (ii) The Hartree term, which does not appear in Leggett approach.
 
As far as the DOS is concerned, we can disentangle two effects, namely the simple fact that at any energy the lattice DOS can be different from the continuum counterpart, and the existence of the cut-off energy $\Lambda$ which directly stems by the existence of a finite lattice spacing.
Even though it is possible to choose the lattice DOS  to be identical to the target DOS at low-energy, as in the case of the {\it{lattice gas}}  models introduced in Sec.  \ref{cut-off}, there is no way to avoid the breakdown of Galilean invariance. 

The Hartree term, absent in the Leggett approach, vanishes as $n^{1/3}$ once rescaled with $E_F$.
Even if it can be easily reabsorbed in a shift of the chemical potential, it plays a leading role in the 
asymptotic corrections to the dilute limit, as we will clarify in the next section . 

Finally we notice that, keeping the Fermi energy $E_F$ fixed (i.e. for fixed density $n$), the same asymptotic equations are achieved if the cut-off $\Lambda$ is divergent, corresponding to a vanishing lattice spacing. This alternative point of view is the analogous of taking the zero-range limit of a continuum model instead of reducing the density of a system with a finite range $b$.

\section{Results}
\label{sec:results}
\subsection{Unitary Limit}
\label{unitarity}

As briefly mentioned in the Introduction, dilute Fermi gases evolve continuously as the interaction is tuned from the BCS limit ($\eta \to -\infty$) to the BEC limit  ($\eta \to \infty$) and a crossover takes place in the region around $\eta=0$. 
We start our analysis of the mean-field results from the unitary limit, where the scattering length diverges, corresponding to $\eta=0$, or equivalently $U=U_c$. 

This limit has attracted an enormous interest because the divergence of $a_s$ makes any length scale other than $k_F^{-1}$ irrelevant leading to an extra universality. Precisely for this reason, most of studies attempting to extrapolate the properties of dilute gases from a lattice perspective have concentrated in this limit.
\begin{figure}[b]
\begin{center}
\includegraphics[scale=0.31]{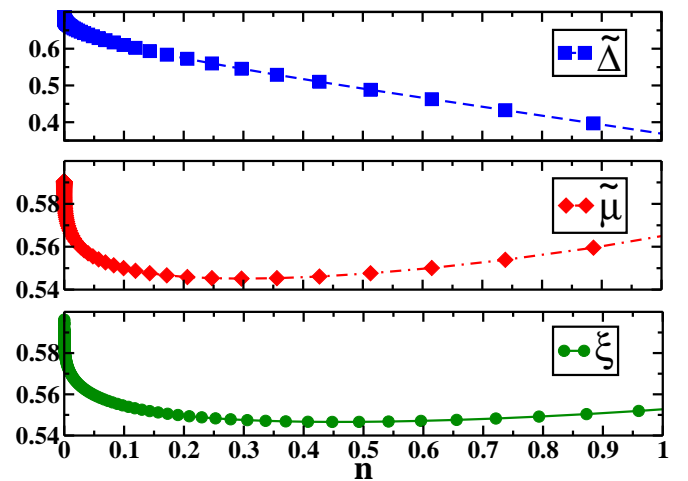}
\caption{(Color Online) Rescaled chemical potential $\tilde{\mu}$, gap $\tilde{\Delta}$, and ground state energy $\xi$  
for the semicircular DOS case along the unitary trajectory ($\eta=0$) as a function of the density $n$.}
\label{fig:mf_unitary_1}
\end{center}
\end{figure}

For the sake of definiteness we start with the results for the semicircular DOS, which presents some advantage in the numerical solutions. In Fig. \ref{fig:mf_unitary_1} we show the rescaled gap parameter $\tilde{\Delta}$, chemical potential $\tilde{\mu}$ and internal energy $\xi=\frac{E}{3/5nE_F}$ as a function of the lattice density $n$. All the quantities correctly converge to Leggett's mean-field values for the dilute gas ($\tilde{\Delta}_L \simeq 0.68$ and $\tilde{\mu}_L= \xi_L \simeq 0.59$) as $n \to 0$\cite{unitary}. Two important facts emerge: All the curves approach their limiting values from below with a divergent slope, and the finite-density corrections are hardly negligible unless  the density is extremely small. 
For example, in order to have a deviation smaller than 1\% for the three observables, we need $n \lesssim 10^{-3}$.

 A free fit of the data in the low-density region $(n \leq 10^{-3})$ shows that the leading
asymptotic corrections to all the observables under consideration are proportional to $n^{1/3}$ for any lattice studied and in the whole BCS-BEC crossover (i.e. for any $\eta$). The same behavior has been reported for the critical temperature in Refs. \cite{burovski1, burovski2, burovski3,goulko}. In Table \ref{tab:summary} we report the coefficients of the $n^{1/3}$ for the three observables $a_{\mu}$, $a_{\Delta}$, and $a_{\xi}$.

\begin{figure}[bht]
\begin{center}
\includegraphics[scale=0.34]{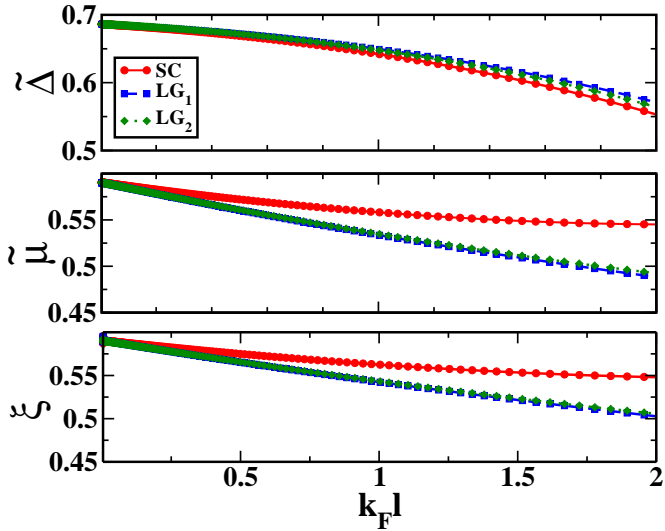}
\caption{(Color Online) Rescaled chemical potential $\tilde{\mu}$, gap $\tilde{\Delta}$ and ground state energy $\xi$ at unitarity ($\eta=0$) as a function of the lattice dilution parameter $k_F l$ for the semicircular DOS (red circle), the lattice gas 1 (blue squares), and lattice gas 2 case.}
\label{fig:mf_unitary_kfl}
\end{center}
\end{figure}

The observed scaling is consistent with the hypothesis that the rescaled observables are analytic functions of $k_Fl$ for every value of $\eta$. The linear term in $\eta$ clearly results in a  $n^{1/3}$ correction at small densities. In Fig. \ref{fig:mf_unitary_kfl} we plot the same quantities as a function of $k_Fl$ for the three lattice models defined above. Even if the corrections to the universal dilute limit are of the same order of magnitude for any lattice, the semicircular DOS gives smaller corrections in $\tilde{\mu}$
and $\xi$ within our mean-field approach.

In the following we discuss the different contributions to these leading order corrections and the actual role of the properties of each lattice. 

The simplest term to analyze is the Hartree correction to the rescaled chemical potential, which at unitarity ($U\equiv U_c$) is simply $\delta \tilde{\mu}=\frac{U_c n}{2 E_F}$ and contributes to the leading-order corrections proportional to $n^{1/3}$.
The numerical value is $\delta \tilde{\mu}=\frac{U_c n}{2 E_F} = -0.171 (k_Fl)$ for the LG$_1$ and it is of the same order for all the lattices. On the other hand  the calculated coefficient $a_{\mu}$ is negative, but significantly smaller in magnitude (see Table  \ref{tab:summary}), suggesting that  the large negative contribution from the Hartree term has to be compensated by a positive contribution which necessarily arises from the difference between the DOS of the lattice and continuum models.

In the following we discuss the origin of these lattice-dependent positive contribution, disentangling a contribution arising from the presence of the cut-off $\Lambda$ and a more subtle contribution due to the fact that the lattice DOS may be different from the continuum one at finite energy.

In order to disentangle these contributions, we consider the $LG_1$ model, in which the DOS is identical to the free one for all energies below the cut-off, which is obviously the only source of corrections.

We can rewrite the mean-field equations (\ref{mf1}) and (\ref{mf2}) for the $LG_1$ model in a compact form as $4/3=A^\Lambdatilde (\mutilde^\prime,\Deltatilde)$ and $\eta=B^\Lambdatilde (\mutilde^\prime,\Deltatilde)$, where 
we introduced
\begin{eqnarray}
\hspace{-0.2cm} A^\Lambdatilde(\mutilde,\Deltatilde) &=& \int_0^\Lambdatilde dx \sqrt{x} \left[ 1-\frac{x-\mutilde}{\sqrt{(x-\mutilde)^2+\Deltatilde^2}}\right] \label{a_lambda}\\
\hspace{-0.2cm} B^\Lambdatilde(\mutilde,\Deltatilde) &=& \frac{1}{\pi}\int_0^\Lambdatilde \hspace{-0.2cm} dx \sqrt{x} \left[ \frac{1}{x}-\frac{1}{\sqrt{
(x-\mutilde)^2+\Deltatilde^2}}\right] \label{b_lambda}
\end{eqnarray}
and $\Lambdatilde =\frac{\Lambda_{LG_1} }{E_F}$. For $n \to 0$ $\Lambdatilde$ diverges and Leggett's equations are recovered. By expanding the equations above for large values of $\Lambdatilde$ and linearizing around the Leggett solution ($\mu_L$, $\Delta_L$), we can derive an analytic expression for the leading-order lattice corrections, i.e.,
\begin{eqnarray}
\mutilde \approx    && \mutilde_L + \frac{1}{(3\pi^2)^{\frac{1}{3}}}\left( \sqrt{\frac{\sigma}{\Lambda_{LG_1} }}C_1 + \frac{U^{LG_1} _c}{2\sigma}\right) k_Fl \label{asympt1}\\
\Deltatilde \approx && \Deltatilde_L+ \frac{1}{(3\pi^2)^{\frac{1}{3}}} \left( \sqrt{\frac{\sigma}{\Lambda_{LG_1} }}C_2 \right) k_Fl ,
\label{asympt2}
\end{eqnarray} 
where $\sigma = E_F/n^{2/3}$ and $C_1$ and $C_2$ are constants calculated in the zero-density limit which do not depend on the lattice 
\begin{eqnarray}
 C_1&&= \frac{
\Deltatilde_L^2  \left(\frac{\partial B}{\partial \Delta}\right)_L 
+ \frac{2\mutilde_L}{\pi} \left(\frac{\partial A}{\partial \Delta}\right)_L
}{
Det[\hat{J}]_L} \\
C_2&&= \frac{
-\Deltatilde_L^2  \left(\frac{\partial B}{\partial \mu}\right)_L 
- \frac{2\mutilde_L}{\pi} \left(\frac{\partial A}{\partial \mu}\right)_L
}{
Det[\hat{J}]_L}
\end{eqnarray}
and $Det[\hat{J}]_L$ is the determinant of the Jacobian matrix $\hat{J}= 
\frac{\partial(A,B)}{\partial (\mu,\Delta)}$ evaluated for $\mutilde=\mutilde_L$
and $\Deltatilde=\Deltatilde_L$.  At unitarity $C_1 \simeq 0.4170$ and $C_2 \simeq -0.0775$.
%0.41695 -0.07751
Eqs. (\ref{asympt1}) and (\ref{asympt2}) are exact expressions for the leading-order density corrections arising from the very presence of a finite cut-off energy in the $LG_1$ case. In the case of the chemical potential this is the term that partially compensates the Hartree shift, while it gives the full leading-order correction to the order parameter. 

These results unambiguously prove that the presence of a cut-off determines $n^{1/3}$ corrections to the observables, 
and there are no self-evident choices of the lattice that can cancel these corrections that ultimately arise from the broken Galilean invariance. 

Obviously for the other lattices, other corrections due to differences between the free DOS and the specific lattice DOS are expected. Indeed the expressions (\ref{asympt1}) and (\ref{asympt2}) (evaluated introducing the proper values of $U_c$ and $\Lambda$ shown in Table \ref{tab:summary}) do not reproduce the calculated coefficients of the $n^{1/3}$ corrections for the semicircular DOS and $LG_1$ models. This implies that also the behavior of the DOS at finite energy contributes to the same order of the cut-off  (and of the Hartree shift in the case of the chemical potential) and therefore leads to further $n^{1/3}$ corrections.

Werner and Castin \cite{castin_long} have shown for a gas in the continuum space that  the leading-order density to the dilute limit assume a universal behavior in terms of the effective dilution parameter $k_F R_e$, where the effective range $R_e$ is defined through the low-momentum $k$ expansion of the scattering amplitude $f$ \cite{castin_varenna}, i.e.
\begin{equation}
f \approx -\frac{1}{a_s^{-1}+ik-R_ek^2/2+\cdots} 
 \label{lowenergyscatt}
\end{equation}

\begin{figure}[!ht]
\begin{center}
\includegraphics[scale=0.325]{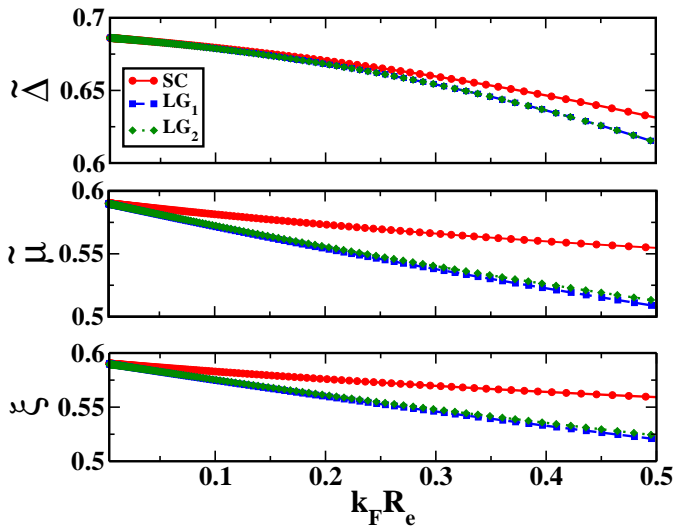}
\caption{(Color Online) Rescaled chemical potential $\tilde{\mu}$, gap $\tilde{\Delta}$ and ground state energy $\xi$ at unitarity ($\eta=0$) as a function of the lattice effective dilution parameter $k_F R_e$ for the semicircular DOS (red circle), the lattice gas 1 (blue squares), and lattice gas 2 case.}
\label{fig:mf_unitary_kfre}
\end{center}
\end{figure}

One could expect a similar universal behavior on a lattice if the asymptotic corrections to the dilute limit  were determined by  the low-energy scattering properties alone. In Fig. \ref{fig:mf_unitary_kfre} we plot the usual observables as a function of $k_FR_e$ using the calculated values of $R_e$ for our lattices.  Our results do not show any universality, confirming that the broken Galilean
invariance, i.e.  the simple existence of a cut-off in the dispersion, plays a key role in the leading-order corrections, which are not determined only by the low-energy scattering properties. In particular within MF the leading order corrections are of order $n^{1/3}$  also for the lattice introduced in Ref. \cite{carlson2011} and designed in order to have $R_e = 0$ (not shown).

Our results also show that there is no easy way to eliminate or systematically the non-universal corrections to the dilute gas behavior even at very small density. The choice of the lattice can be important only to quantitatively  reduce the finite-density correction, but no lattice can remove the effects of the reduced translational invariance.

\subsection{Crossover Region}
\begin{figure}[hb]
\begin{center}
    \begin{tabular}{c}
     \resizebox{90mm}{!}{\includegraphics[angle=-90]{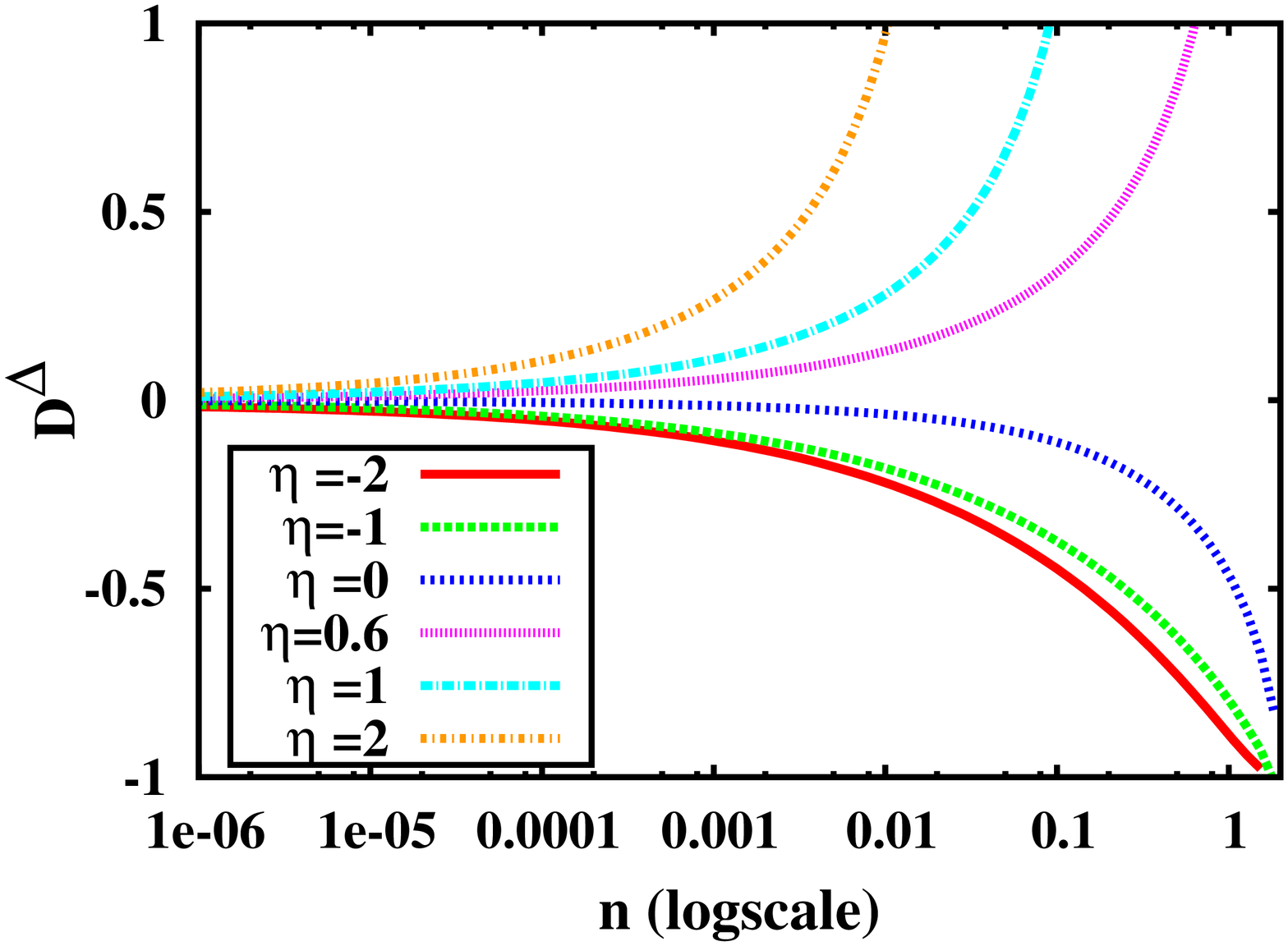}} \\
     \resizebox{90mm}{!}{\includegraphics[angle=-90]{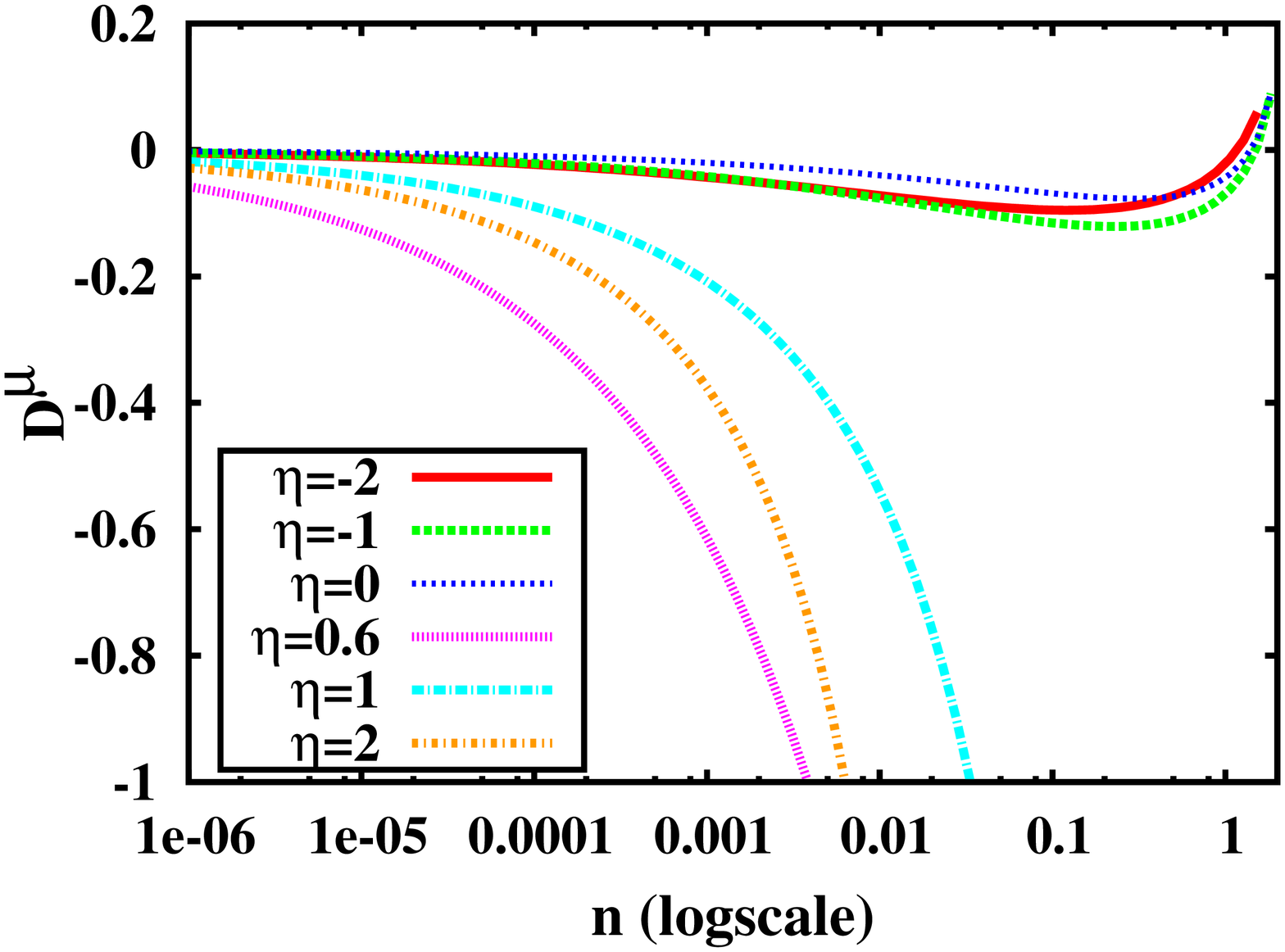}} 
    \end{tabular}
\end{center}
\caption{(Color online)  Relative deviation $D^\Delta$ (upper panel) and $D^\mu$ (lower panel) 
of the rescaled parameters from Leggett values 
as a function of the density (logscale) for different values of the gas parameter $\eta $ 
($\eta =-2,-1,0,0.6,1,2$). }
\label{fig:fig7} 
\end{figure} 
  
In this section we extend our analysis beyond the unitary point. Therefore we solved the mean-field equations as a function of the density along the trajectories at constant $\eta=$ in the whole crossover spanning the region $-2 < \eta < 2$. We start from an overview of the results for the SC DOS. 

For every value of $\eta$ we correctly recover Leggett's results in the dilute limit.
Since the limit values $\tilde{\Delta}_L $ and $\tilde{\mu}_L $ strongly depend on $\eta$ and vary over several orders of magnitude in the crossover, it is more instructive to plot the relative deviations $D$ (upper panel of Fig. \ref{fig:fig7}) and analogously $D^\mu$ (lower panel) as a function of the density for several values of $\eta$. 
     
A first general conclusion is that the approach to the asymptotic values is slow for all values of $\eta$, confirming that, at any value of the effective interaction, a reliable extrapolation to the dilute limit requires a reasonably small density, whose value depends however on $\eta $. 

As expected from the general considerations in Sec. \ref{sec:mapping}, the slowest convergence is observed on the BEC side $\eta>0$
where the corrections are much larger than for the unitary point. The large values of $D^\Delta$ observed deep in the BCS regime are essentially due to a vanishingly small value of the order parameter,  and the same apply for $D^\mu$ when $\eta=0.6$ since the chemical potential changes sign close to this point.

The gap parameter $\Delta$ is in general less sensitive to finite-density effects than the chemical potential.  In the asymptotic low-density region, the deviation $D^\Delta(n)$ is a monotonically increasing function of $\eta$ for fixed $n$. The gap parameter is overestimated coming from the BEC side of the crossover and underestimated coming from the BCS side, even though the gap corrections along the unitary trajectory $\eta=0$ are more similar to the BCS region, i.e., they converge from below.
In the same range of densities, $D^\mu$ is negative  for every value of $\eta$ we investigated, i.e., finite density reduces the rescaled chemical potential at any $\eta$.
\begin{figure}[b]
  \begin{center}
  \includegraphics[scale=0.325]{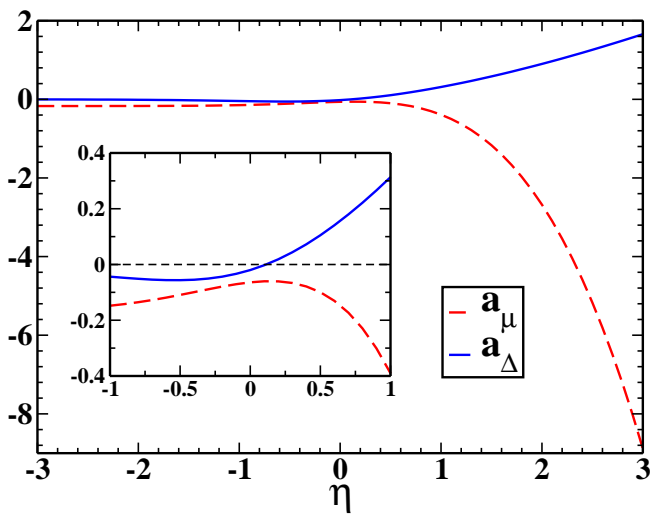}
  \caption{(Color Online) Asymptotic slope of the lattice corrections to the chemical potential $a_\mu^{LG_1}$ and to the gap parameter $a_\Delta^{LG_1}$ for the $LG_1$ model as a function of the inverse gap parameter in the BCS-BEC crossover. The inset show a magnified view 
of the the region around the unitary point.}
   \label{fig:Fig8}
 \end{center}
 \end{figure}  

We can extend our analysis of the leading-order corrections to the dilute limit. We start from the Hartree shift of the chemical potential
$\delta \tilde{\mu}= \frac{-n|U_\eta(n)|}{2 E_F}$
where we emphasized the dependence of $|U|$ on $n$ along the trajectories at constant $\eta$. Since $U_\eta(n)$ differs from $U_c$ only at the order $n^{1/3}$ (See Eq. (\ref{U_eta})) the leading-order correction ($\propto n^\frac{1}{3}$) due to this term is identical to the one at unitarity for every value of $\eta$. 
 
This observation implies that the analytic expressions for the leading-order lattice corrections due to the cut-off in the $LG_1$ model given by Eqs. (\ref{asympt1}) and (\ref{asympt2}) are valid in the whole crossover, provided that we evaluate $C_1$ and $C_2$ using $\mutilde_L(\eta)$ and $\Deltatilde_L(\eta)$ which solve the Leggett equations for the given $\eta$. In Fig. \ref{fig:Fig8},  we show the results for $a_{\mu}(\eta)$ and $a_{\Delta}(\eta)$ in the crossover region. 

In the BCS limit $\eta \to -\infty$ the contribution due to the cut-off is vanishingly small,  $C_1 \to 0^+$ and $C_2\to 0^-$. Therefore $a_{\mu} \to -0.171$, given by the Hartree shift, and $a_{\Delta} \to 0^-$. The vanishing of the correction due to the cut-off can be understood by considering that in the BCS limit pairing only involves an exponentially small region of energies around the Fermi energy $E_F \ll \Lambda$ and the cut-off region gives an exponentially small contribution.

In the case of the $LG_1$ model the Hartree shift and the cut-off contribution alone determine the leading-order correction, but on a generic lattice, like, e.g., the semicircular DOS model, we can have contributions arising from the differences in the functional form of the density of states.

Computing explicitly these corrections for the semicircular DOS in the BCS regime, we find that they are of order $(k_F l)^2\propto n^{2/3}$.  This observation shows that at weak-coupling, the difference in the DOS does not contribute to the leading-order corrections. On the other hand, as shown in our results about the unitary point, this is not true outside the BCS regime. 

Turning back to the $LG_1$ model, we find that beyond the BCS regime $C_1$ and $C_2$ are always sizeable, generalizing the result obtained at unitarity, that the cut-off plays a crucial role in determining the leading-order corrections even if $\Lambda \gg E_F$. This is essentially due to fact that outside the BCS regime pairing involves a wider region of energies whose size increases going to the BEC limit. As expected, in the BEC regime the size of the corrections explodes,  being several order of magnitude larger already for moderately large values of $\eta$. 
As evident from the inset in Fig. \ref{fig:Fig8}, $a_{\mu}$ is always negative  and has a rather broad maximum slightly shifted towards the BEC side. $a_{\Delta}$ always increases with $\eta$ being negative in the BCS limit and positive in the BEC limit. It is interesting to notice that the unitary point $\eta=0$ does not show any special behavior with respect to the finite-density lattice corrections. 

\section{outlook}
Our analysis shows that the presence of a lattice is hardly irrelevant even at very small density. Moreover we have shown that there is no easy shortcut to overcome this poor convergence as a function of density. Even what we label lattice gas 1, whose DOS differs from the bare one only for the unavoidable presence of a cut-off, has sizeable leading-order corrections, and the reduction of the effective range can only reduce the quantitative effects.

These considerations are crucial in light of the explosion of lattice approaches to BCS-BEC crossover in dilute Fermi gases, mainly based on QMC simulations \cite{bulgac1,bulgac2,bulgac3,burovski1,burovski2,burovski3,goulko,lee1, lee2,drut} and DMFT \cite{barnea}.
We remark that our analysis only refers to lattice simulations and it does not pretend to predict the behavior of low-density corrections in continuum space.
 
QMC approaches for attractive models provide very accurate results (or at the very least upper bounds) for {\it finite-size} lattice models {\it at finite density $n$}. Our point is that, even for exact methods, a great care is needed to extract reliable information on the actual dilute system in the thermodynamic limit.
 
Even after a careful extrapolation to the thermodynamic limit is performed in order to eliminate finite-size effects,  a further extrapolation of the dilute limit $n \to 0$ is still required to access universal and therefore model-independent features in the dilute regime.

Precise results require therefore simultaneously large lattices and large number of particles. We argue that one of the reasons why current QMC results are scattered around the most precise experimental estimates (see e.g. \cite{nascimbene,navon,horikoshi,zwierlein}) rather than converging to them, is that non-universal lattice features are still present at the densities considered in these approach and not properly taken into account. 

Our analysis clearly shows that choosing a given density and assuming that it is small enough for the properties of the system to coincide with the dilute limit can lead to substantial deviations, especially if the densities are of order $ n \sim 0.1$ as in Refs. \cite{bulgac1,bulgac2,bulgac3}.
At these densities the deviations from the dilute gas behavior can not be considered as small (see e.g. Figs \ref{fig:mf_unitary_1} and \ref{fig:fig7}), especially if the aim is the quantitative evaluation of the dilute gas properties, like, e.g., the existence of a small pseudogap at unitarity. The agreement with recent experimental results \cite{nascimbene} is most likely resulting from a cancellation  of residual finite-size and finite-density effects.

While the need of a careful extrapolation to the dilute limit has been realized by different authors (see e.g. \cite{burovski2,goulko}), the lack of an underlying theory which explains the actual relevance of these corrections  makes the extrapolation potentially uncontrolled. In particular, it is important to understand that one does not simply need to reach small densities, but it is also necessary to include in the fit {\it only} data in a region close to $n=0$ where a proper power-law behavior is present. Our investigation shows that this region can be very small, and densities of the order ot $n \sim 0.1$ are definitely outside and should not be included in the fit.

For example, the different values of the critical temperature at unitarity calculated in Ref. \cite{goulko} and Ref. \cite{burovski1} are likely explained by the lower minimal density used in Ref. \cite{goulko}, $n_{min} \approx 0.01$, five times smaller than  $n_{min} \approx 0.05$ in Ref. \cite{burovski1}, even if a different size extrapolation has been invoked to justify the discrepancy. 

We also notice that the linear fit to extrapolate $T_c$ at unitarity in Ref. \cite{goulko} includes data at density as high as $n \approx 0.3$, which are clearly not in the asymptotic $n^\frac{1}{3}$  regime according to the results in this paper and in Ref. \cite{ufg_old}. Also this effect is likely to affect the extrapolation and, as a general remark, only the data at the lowest should be used in the fit procedures in order to extract precise estimates for the dilute Fermi gas free from uncontrolled errors.  

 \section{Conclusions}
 \label{sec:conclusions}
In this paper we studied the connection between the BCS-BEC crossover in dilute gases and in lattice models at finite density.
More specifically  we investigated within mean-field how several relevant observables in different lattice models all converge to those of a dilute gas as the density is reduced along suitable trajectories with constant gas parameter $k_F a_s$.   Comparing with the solution of a dilute Fermi gas in continuum space, we analyzed quantitatively how the properties of lattice systems converge to the dilute limit.

In analogy with the finite-size effects, which can be minimized through the knowledge of the expected dependence on the size of the system, a theoretical understanding of the power-law density corrections to the dilute regime is a crucial step towards a proper zero-density extrapolation.
The mean-field results presented in this paper, though approximate, are meant to be a guideline to control the extrapolation. Moreover, mean-field results are free from the interplay with finite-size effects, since our approach directly deal with the problem in the thermodynamic limit. 

Our results clearly show that, despite the universal nature of the dilute limit, finite density corrections depend on the specific choice of lattice. Moreover, these corrections persist down to extremely low densities, making any extrapolation based on lattice model at finite density very hard unless data at extremely low-densities are available. The leading order corrections to all the relevant observables at small densities follow a $n^{1/3}$ power-law behavior for every lattice studied  here and in Ref. \cite{ufg_old}. The prefactors instead depend on the specific lattice and vary over several order of magnitude as the lattice parameter spans the BCS-BEC crossover, being larger on the BEC side. 
The $n^{1/3}$ behavior also holds for a model with the same DOS of free fermions except for a sharp energy-cut-off which enforces the reduced translational symmetry of the lattice. In this case we also derived an exact analytical expression for the leading-order prefactors in the whole crossover.

This result shows unambiguously that the presence of a high-energy cut-off $\Lambda$ associated with the broken Galilean invariance affects the properties of the system down to arbitrarily low finite densities. This finding  contradicts the common belief -valid only in the weak-coupling BCS regime- that only the low-energy behavior up to the Fermi energy $E_F \propto n^{2/3} \ll \Lambda$ determines the leading-order correction to the dilute limit. 
Comparing different lattices we show that also the differences of the DOS at finite energy far from the Fermi level contribute to the leading-order corrections. As  a consequence, the prefactors of the leading density corrections are non universal and there is no simple recipe to cancel or even minimize them. 

The relevance of the DOS contrasts the idea that the leading-order corrections depend only on the low-energy scattering properties. This rules out the possibility to cancel the leading-order density corrections tuning the effective range to zero as proposed in \cite{carlson2011}, and seems to exclude a simple way to express the corrections in terms of a single parameter.

The inclusions of fluctuations beyond mean-field is crucial for a quantitative description of the BCS-BEC crossover both in the dilute gas and in lattice models. However, fluctuations can even increase the sensitivity of the system to density variations, as found in Ref. \cite{ufg_old} comparing static and dynamical mean-field theory for lattice models at unitarity and in \cite{cinese} where the relevance of density corrections to extrapolate the critical temperature is investigated including Gaussian fluctuations.  The same enhancement of lattice-dependent features is expected to happen in finite-temperature data with respect to $T=0$ \cite{lee_finiteT,cinese}.

During completion of the present work, we became aware of a very recent Monte Carlo study\cite{newqmc} which shows a sizable effect of reducing the fermionic density from $n \simeq 0.1$ to $n \simeq 0.04$, in line with our results.

\begin{acknowledgements}
We are indebted with C. Castellani for precious discussions in the early stages of this work. AP thanks C. Verdozzi and the Divison of Mathematical Physics at Lund University for their warm hospitality during the completion of this work.
We acknowledge financial support by European Research Council under FP7/ERC Starting Independent Research Grant ``SUPERBAD" (Grant Agreement n. 240524).
\end{acknowledgements}

\end{document}